\documentclass[twocolumn,preprintnumbers,amsmath,amssymb,superscriptaddress]{revtex4}
\usepackage{graphicx}
\usepackage{dcolumn}
\usepackage{bm}
\usepackage{natbib}

\def\ltsima{$\; \buildrel < \over \sim \;$}
\def\simlt{\lower.5ex\hbox{\ltsima}}
\def\gtsima{$\; \buildrel > \over \sim \;$}
\def\simgt{\lower.5ex\hbox{\gtsima}}

\def\[{\begin{equation}}
\def\]{\end{equation}}
\def\m@th{\mathsurround=0pt }
\def\eqalign#1{\null\,\vcenter{\openup1\jot \m@th
 \ialign{\strut\hfil$\displaystyle{##}$&$\displaystyle{{}##}$\hfil
 \crcr#1\crcr}}\,}

\begin{document}
\title{Redshift Sensitivity of the Kaiser Effect}

\date{\today}
\newcommand{\ud}{\mathrm{d}}
\newcommand{\fpe}{f_\perp}
\newcommand{\fpa}{f_\parallel}
\newcommand{\om}{\Omega_m}

\author{Fergus Simpson}
 \email{frgs@roe.ac.uk}
\affiliation{SUPA, Institute for Astronomy, University of
Edinburgh, Royal Observatory, Blackford Hill, Edinburgh EH9 3HJ}

\date{\today}

\begin{abstract}
We explore potential strategies for testing General Relativity via the coherent motions of galaxies. Our position at $z=0$ provides the reference point for distance measures in cosmology. By contrast, the Cosmic Microwave Background at $z \simeq 1100$ acts as the point of reference for the growth of large scale structure. As a result, we find there is a lack of synergy between growth and distance measures. We show that when measuring the gravitational growth index $\gamma$ using redshift-space distortions, typically $80\%$ of the signal corresponds to the local growth rate at the galaxy bin location, while the remaining fraction is determined by its behaviour at higher redshifts.

In order to clarify whether modified gravity may be responsible for the dark energy phenomenon, the aim is to search for a modification to the growth of structure. One might expect the magnitude of this deviation to be commensurate with the apparent dark energy density $\Omega_\Lambda(z)$. This provides an incentive to study redshift-space distortions (RSD) at as \emph{low} a redshift as is practical. Specifically, we find the region around $z = 0.5$ offers the optimal balance of available volume and signal strength.

\end{abstract}
\maketitle

\section{Introduction}

Dark energy is a low-redshift phenomenon, and in accordance with the standard $\Lambda CDM$ model, it appears to exert a rapidly decaying influence towards higher redshifts, at a rate approaching $(1+z)^{-4}$. Physical models of dark energy may be distinguished by an equation of state $w \neq -1$, while a break from general relativity would likely exhibit a distinctive structure formation history. The consequences of a modification to general relativity are somewhat speculative at this stage, but naturally one might expect any alteration of the growth rate to become particularly prominent at late times, in accordance with the observed change in global dynamics. For instance, the $f(R)$ models explored by Hu \& Sawicki \cite{2007PhRvD..76j4043H} found this to be the case on large scales, as did He et al \cite{2009JCAP...07..030H} with their interacting model.

A number of probes are capable of measuring $w$ via its influence
on the redshift-distance relation. It is rather more difficult to
study the growth rate, due to our uncertainty in the behaviour of galaxy
bias, but there are currently two promising avenues available for future
exploration. Weak gravitational lensing provides a direct measurement
of the dark matter distribution, and its evolution with redshift. However the focus of this work will be redshift-space distortions, which exploit
the relationship between the large-scale coherent velocities of galaxies
and the growth rate of perturbations.

Weight functions have previously been applied to the equation of state as a function of redshift $w(z)$ \cite{2003MNRAS.343..533D,2005PhRvD..71h3501S,2006PhRvD..73h3001S}, and this work extends the concept to the growth index $\gamma$ \cite{2005PhRvD..72d3529L}. They are designed to illustrate the redshift sensitivity of a given survey, and this is often quite different from the source redshift distribution. Weight functions are closely related to, and may be derived from, the principal component approach. In related work, Zhao et al. \cite{2009arXiv0905.1326Z} recently explored the principal components of the metric ratio, a different quantity also often denoted by $\gamma$. 

In \S\ref{sec:wts} we briefly review the concept of principal components and their application in deriving the weight function. \S\ref{sec:rsd} explores the RSD redshift sensitivity to the growth index $\gamma$, and compares these weight functions to those of the dark energy equation of state. \S\ref{sec:opt} addresses the question of which redshift is most efficient at measuring $\gamma$, while in \S\ref{sec:gro} we consider a change of parameterisation to the growth rate $f$.

\section{Weight Functions} \label{sec:wts}

The main purpose of a ``weight function" is to illustrate the redshift range over which we are most sensitive. This has previously been applied to the dark energy equation of state \cite{2003MNRAS.343..533D,2005PhRvD..71h3501S,2006PhRvD..73h3001S}, where the best-fit constant equation of state $w^{fit}$ is expressed as a weighted integral over the true functional form

\[
w^{fit} = \int \Phi_{w}(z) w(z) \ud z   .
\]

\noindent Here we apply this approach to the growth index $\gamma$ \cite{2005PhRvD..72d3529L} (not to be confused with another modified gravity parameter, the metric ratio), defined by the approximate relation \cite{1998ApJ...508..483W}

\[
f(z)\equiv\frac{\ud\ln\delta}{\ud\ln a}\simeq\Omega_{m}^{\gamma}(z)  ,\]

\noindent where $f(z)$ is the logarithmic growth rate of linear perturbations. Ordinarily this is well approximated by $\Omega_m^{0.55}$, but in order to act as a phenomenological test of gravity the exponent is treated as a variable, denoted by $\gamma$. Much like the equation of state $w(z)$, an important aim for future cosmological surveys will be to test whether the growth index $\gamma$ matches the value corresponding to that expected from General Relativity, despite the time-dependence of any deviation remaining highly uncertain.  Few feasible alternatives to GR are yet known, but DGP \cite{2000PhLB..485..208D} is one example where the growth index has been evaluated, with $\gamma_{DGP} = 0.69$ \cite{2007APh....28..481L}. Due to its scale-dependent growth factor, f(R) models are not so well described by this simple parameterisation. However it is worth noting that in general these models enhance linear growth and as such typically exhibit lower values at around $\gamma \sim 0.4$ \cite{2009PhRvD..80h4044T}.

The main focus of this work is to identify the redshift at which we are observing $\gamma$. We reach a quantitative solution by considering the constant value of $\gamma$ one would infer from an arbitrary function $\gamma(z)$, as given by  

\[ \label{eq:gammawt}
\gamma^{fit} = \int \Phi_{\gamma}(z) \gamma(z) \ud z  .
\]

\noindent 

In order to generate $\Phi(z)$, we begin by decomposing the function $\gamma(z)$ into orthonormal eigenfunctions $e_i(z)$ as given by

\[
\gamma(z) = \sum_i \alpha_i e_i (z)  .
\]

\noindent As shown in \cite{2006PhRvD..73h3001S}, the weight function may be expressed as a sum of the eigenmodes $e_i (z)$ weighted by the errors $\sigma_i$ associated with the eigenvalues $\alpha_i$, along with a second weighting term which penalises highly oscillatory modes, for they will contribute little when fitting a constant value.

\[
\label{eq:phipca} \Phi(z)= \frac{\sum_i{e_i(z)\int e_i(z') \ud
z'/\sigma^2(\alpha_i)}}{\sum_j \left[\int e_j(z'') \ud z''\right]^2
/\sigma^2(\alpha_j)} \,\,.
\]

In general the form of the weight function is invariant to the absolute error of the survey, and for a given technique its strongest dependence lies with the source redshift distribution. 

In recent work by Kitching \& Amara \cite{2009arXiv0905.3383K}, the authors highlight the potential dangers of evaluating the eigenmodes using a finite set of discretised redshift bins, given the variation which may arise when adopting different basis sets. While this may well prove problematic for studies which adopt fewer than $\sim 30$ bins, we ensure adequate convergence by dividing the scale factor into $300$ bins. 

Since we are essentially trying to determine the Green's function associated with the observable, we would ideally use Dirac's $\delta$-functions to perturb the parameter in question, and in some cases such as supernovae this step may be performed analytically \cite{2003MNRAS.343..533D}. However, given that $\gamma$ enters as a raw exponent, as opposed to the integrated form that $w(z)$ appears in, we are restricted to working numerically.

\begin{figure}
\includegraphics[width=80mm]{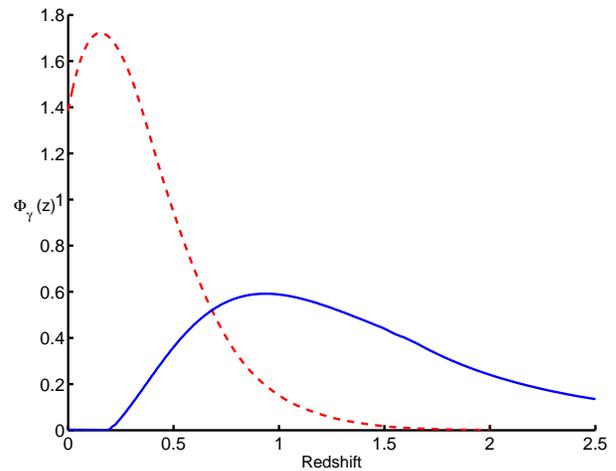} 

\caption{The redshift sensitivity of structure formation with linear redshift-space distortions. The solid line shows the weight function $\Phi_\gamma$, as defined in (\ref{eq:gammawt}), where $\gamma$ is the gravitational growth index.  The survey spans the redshift range $0.2<z<1.8$, with fixed solid angle and constant comoving number density, and is combined with Planck. The dashed line is the weight function for fitting a constant equation of state $\Phi_w$, for a typical high-redshift supernova and weak lensing survey, with sources spanning the same redshift range.}

\label{fig:rsd_wt}
\end{figure}

\begin{figure}
\includegraphics[width=80mm]{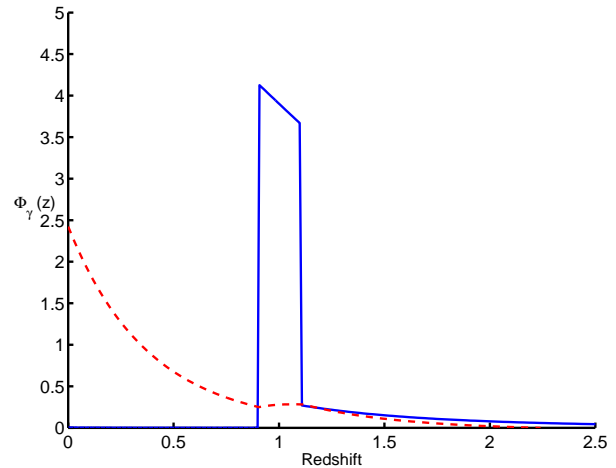}

\caption{A comparison of the weight functions for  $\Phi_\gamma$ (solid) and  $\Phi_w$ (dashed) using RSD and BAO from the same sample of galaxies at a mean redshift $z=1$ and with a bin width $\Delta z = 0.2$. This highlights the restricted range of redshift sensitivity for $\gamma(z)$, in particular it vanishes below the bin edge. There is also very little overlap between the two functions.}

\label{fig:rsd_wt_bin}
\end{figure}

\section{Redshift-Space Distortions} \label{sec:rsd}

In redshift space the galaxy power spectrum is significantly anisotropic, and may be expressed as a function of the line-of-sight and tangential components of the Fourier modes, denoted $k_{\parallel}$ and $k_{\perp}$ respectively.
On sufficiently large scales it is well described by

\[ \label{eq:pkkaiser}
P(k_{\parallel},k_{\perp})=P(k)\left(1+\beta\mu^{2}\right)^{2},\]

\noindent where $\mu=k_\parallel / |\bf{k}|$, and $\beta = f/b$, with the local linear galaxy bias $b$ taken to be scale-independent. Combined with the CMB, the observable quantity is essentially given by \cite{2001MNRAS.327..689T,2009MNRAS.393..297P, simpsonp09}

\[ \label{eq:fsig}
\frac{f(z) \sigma_8(z)}{\sigma_8(z_{CMB})}  ,
\]

\noindent and  $\sigma_8(z)$ is the amplitude of dark matter perturbations at the redshift of the survey, $z$. Whilst this approach does assume that the background cosmology is already known, the form of the weight function is independent of the absolute error associated with the survey, and we find this description generates eigenmodes in very good agreement with those from the more rigorous approach presented in \cite{simpsonp09}. 

Figure \ref{fig:rsd_wt} demonstrates the redshift sensitivity which is achieved when studying the rate of structure formation with linear redshift-space distortions. Compared to that of the equation of state, a tendency towards substantially higher redshifts should not be too surprising given that the benchmark of structure growth is the cosmic microwave background. Any geometrical measure will rely on the behaviour of $w(z)$ all the way down to $z=0$, for a given $\Omega_{m}(z=0)$. 

The difference becomes more striking when in Figure \ref{fig:rsd_wt_bin} we compare these two sensitivities for sources in a limited redshift range $0.9<z<1.1$. A spectroscopic survey which generates will inevitably produce a simultaneous measure of the BAO at the same redshift. Much like lensing and supernovae, the BAO harbours most of its sensitivity at low redshift $z<0.5$, while the RSD measure $\gamma(z)$ in the bin itself ($\sim 80\%$), with the remaining weight all attributed to higher redshifts. This fraction remains fairly robust to a change in redshift.

The lack of overlap between the two weight functions seen in Figure \ref{fig:rsd_wt_bin} is of concern, since even if a deviation of w=-1 was discovered, and $\gamma$ was in turn found to be perfectly consistent with General Relativity, this would leave us in a position where we are still unable to deduce the nature of the underlying physics. In other words, these observations have two equally plausible explanations - either (a) dark energy takes the form of a physical fluid with $w \neq -1$, within the framework of General Relativity, which recently took control of the cosmological expansion rate, or alternatively (b) cosmological dynamics are governed by a modified theory of gravity, generating an effective equation of state $w_{eff}(z)$ and a modified $\gamma(z)$, but which only became apparent at some lower redshift. Therefore given that the weight function $\Phi_\gamma$ is concentrated exclusively at $z>0.9$, we would be left unable to distinguish between the physics of these very different scenarios.

For redshift-space distortions, the spike in $\Phi_\gamma$ within the redshift bin is associated with the measure of $f(z)$ from \ref{eq:fsig}, whereas the non-zero sensitivity at higher redshift arises through the $\sigma_8(z)$ term, since this is a cumulative effect. It is typically found that almost $80\%$ of the weight lies within the redshift bin. As we shall see in \S\ref{sec:gro}, this is partly due to the nature of our parameterisation.

\begin{figure}
\includegraphics[width=80mm]{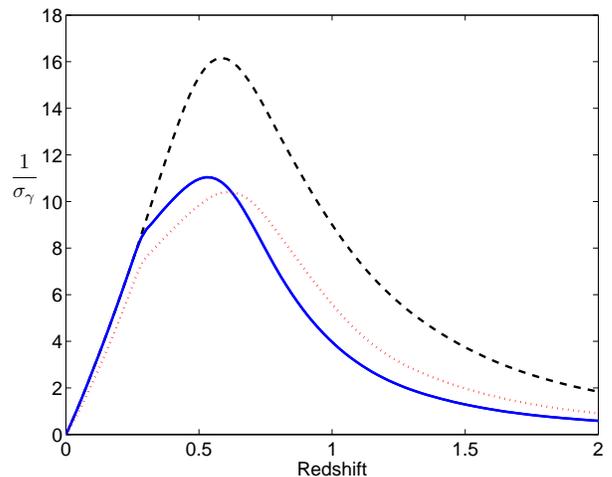}
\caption{The precision with which the parameter $\gamma$ could be measured, as a function of the position of a single redshift bin of width $\Delta z = 0.2$.  The dashed line corresponds to the simple case of a 20,000 square degree survey, with a fixed comoving number density at $n=10^{-3} h^{3}\mathrm{Mpc}^{-3}$, demonstrating the limitations of even the most ambitious high-redshift survey. For the solid line we impose an upper limit on the area density $N = 50 \, \rm{gals \, deg}^{-2}$. The dotted line illustrates the impact of changing the background cosmology such that $w=-0.9$. This leads to more dark energy at high redshift, pushing the peak out to $z \simeq 0.6$. }
\label{fig:dgamma}
\end{figure}

\begin{figure}
\includegraphics[width=80mm]{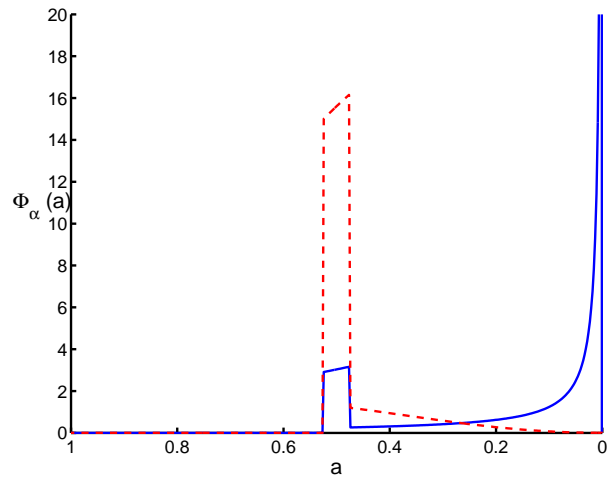}
\caption{The sensitivity of redshift distortions to the modified growth rate $f(z)$, as given by (\ref{eq:phif}) for a redshift bin of width $0.2$ at $z=1$. The dashed line is equivalent to the solid line in Figure \ref{fig:rsd_wt_bin}, but now recast in terms of the scale factor in accordance with (\ref{eq:scalefac}).}
\label{fig:figf}
\end{figure}

\section{Optimal Survey Strategies} \label{sec:opt}

One aspect the weight function does not address is the question of precision. Here we briefly review the impact that the position of a single galaxy redshift bin has on the absolute error on $\gamma$.

Whilst it will undoubtedly be of interest to explore the growth rate of structure across all observable redshifts, the priority must lie with the regime which maximises the opportunity of a significant result. We emphasise that at present dark energy is only known to be a low-redshift phenomenon; its influence at any other epoch remains highly speculative. As such it appears advantageous to study the growth of structure in regions of high $\Omega_\Lambda$, and the parameterisation of $\gamma$ offers a natural mechanism for this preferential weighting.

This raises the question, galaxies at which redshift are best suited to measuring $\gamma$? To help answer this, the plot in Figure \ref{fig:dgamma} illustrates the precision with which $\gamma$ may be determined, as a function of redshift bin position. Rather than the simplified approach of treating $f \sigma_8$ as the observable, in this section we adopt the methodology outlined in the companion paper \cite{simpsonp09}. This essentially results in a simultaneous measure of the baryon acoustic oscillations, redshift space distortions, combined with the DETF Fisher matrix for Planck. To begin we consider the simplest case of the cosmic variance limit, such that the error on $P(k)$ is dictated by the available comoving volume at that redshift. We assume a sky coverage of $20,000$ square degrees, and marginalise over the parameters $[w_0, w_a, \Omega_\Lambda,  \Omega_k, \Omega_m h^2, \Omega_b h^2, n_s, A_s, \beta, \gamma,]$ . However the emphasis here is not on the absolute value of the error itself, but how this changes as a function of the source redshift, which is relatively insensitive to our choice of parameter set.

At very low redshift the dominant factor is cosmic variance, due to the limited comoving volume available. Beyond the optimal peak close to $z=0.5$  there is simply a lack of ``dark energy" at high redshift, which leads to $\Omega_m \simeq 1$ and thus $\sigma_\gamma$ diverges.

\[
\delta\gamma=\frac{1}{\ln\left(\om(z)\right)} \frac{\delta f}{f}  .
\]

A further contributing factor is the diminishing amplitude of the dark matter power spectrum at higher redshifts.

Note that the dashed line in Figure \ref{fig:dgamma} assumes a constant comoving number density, neglecting the inevitable decline towards higher redshifts, and the associated increase in observing time due to the greater volume covered. These factors further disadvantage the prospects of high-redshift studies. Conversely, at low redshifts non-linearities in the power spectrum may prove problematic, as the simple form presented in (\ref{eq:pkkaiser}) will likely prove inadequate. 

At high redshift this corresponds to a prohibitively large number of sources.
The solid line in Figure \ref{fig:dgamma} reflects the error on $\gamma$ achievable with an upper bound on the number of sources per unit area ($N = 50 \, \rm{deg}^{-2}$). Whilst there are a variety of choices and assumptions that can be made regarding the survey parameters, most of these are only found to significantly affect the absolute error, leaving the peak position largely unaffected at $z \sim 0.5$. For the choice of fiducial background cosmology, changes in the abundance of dark energy will have some impact on the form of this function. For example, the dotted line illustrates the influence of setting $w=-0.9$, which corresponds to a greater $\Omega_\Lambda$ at high redshift and thus a stronger signal. Conversely, a value of $w=-1.1$ generates a stronger supression at high redshift. We also find that adding auxiliary information from geometric probes such as supernovae tends to preferentially strengthen the low-redshift measurements, shifting the peak down to $z \sim  0.3$.

\section{Sensitivity to the Growth Rate} \label{sec:gro}

The high-redshift sensitivity diminishes in Figures \ref{fig:rsd_wt} and \ref{fig:rsd_wt_bin} due to the nature of the parameter $\gamma$, such that at high redshift when $\Omega_m$ approaches unity, the value of $\partial f / \partial \gamma$ vanishes.  As a comparison, it is of interest to consider $f(z)$ as the function to perturb instead of $\gamma$. To generate the appropriate weight function, the fiducial growth rate $f(z)$ is modulated by a prefactor $\alpha(z)=1$, such that

\[
f(z) =  \alpha(z) \Omega_m^{0.55}(z)   ,
\]

\[ \label{eq:phif}
\alpha^{fit} = \int \Phi_{\alpha}(z) \alpha(z) \ud z   .
\]

The corresponding weight function $\Phi_\alpha$ is shown in Figure  \ref{fig:figf}. Since a substantial degree of sensitivity now extends up to $z=1100$, it is more appropriate to view this in terms of the scale factor. Given the requirement $|\Phi(z) \ud z| = |\Phi(a) \ud a|$, the rescaling is simply given by

\[ \label{eq:scalefac}
\Phi(a)=\Phi(z)(1+z)^2   .
\]

Since the majority of fractional structure growth occurs at early times, a strong bias towards this epoch can be seen in Figure \ref{fig:figf}.

\section{Conclusions}

At present there are two key approaches to study the evolution of dark matter perturbations, namely redshift-space distortions and weak gravitational lensing. They will provide a crucial piece of evidence in determining whether the phenomenon of dark energy may be attributed to a physical entity, or is simply due to a misunderstanding of the laws of gravity.

In this work we have quantified the epoch at which a galaxy redshift survey would be sensitive to the growth index $\gamma$. Specifically, given the absence of any weight at redshifts lower than that of the survey, low redshift surveys are left with the significant advantage of probing a broader behaviour $\gamma(z)$. Another interesting feature we have highlighted is that approximately $80 \%$ of the ``weight" for $\gamma$ for any given redshift bin corresponds to the local value of $\gamma(z)$ at the location of the bin.

The main limitation of a low redshift survey would be the available comoving volume, along with the stronger prevalence of nonlinear perturbations which may hinder an accurate determination of $\beta$. However, for a given error on $\beta$, one reaches a much better determination of $\gamma$ compared to higher redshifts. 

By contrast, a weak lensing survey with source galaxies at any redshift will still be sensitive to the value of $\gamma(z)$ across \emph{all} redshifts $(0<z<1100)$. Although this may be restricted at very low redshift by the lensing efficiency.

For a purely cosmic-variance limited redshift survey, it appears the shell surrounding $z \sim 0.5$ is optimal in its efficiency. Incorporating more realistic constraints, such as limited observing time, will likely serve to lower this value further unless non-linearities prove highly problematic. This preference towards a low-redshift measurement reflects the increased likelihood of discovering a modified gravity signature when searching within the era of dark energy.


\noindent{\bf Acknowledgements} \\
The author would like to thank S. Bridle, A. Heavens, J. Peacock and P. Norberg  for helpful comments. FS is supported by an STFC rolling grant, and acknowledges the hospitality of the Aspen Center for Physics where part of this work was undertaken.

\bibliography{M:/Routines/dis}

\end{document}